\documentclass[12pt]{article}
\usepackage{geometry}
\usepackage{authblk}
\usepackage{url}
\usepackage{graphicx}%
\usepackage{multirow}%
\usepackage{amsmath,amssymb,amsfonts}%
\usepackage{amsthm}%
\usepackage{mathrsfs}%
\usepackage[title]{appendix}%
\usepackage{xcolor}%
\usepackage{textcomp}%
\usepackage{manyfoot}%
\usepackage{booktabs}%
\usepackage{listings}%
\usepackage{url}
\usepackage{bm}
\usepackage{algorithm}
\usepackage[noend]{algorithmic}
\usepackage{tabularx}
\usepackage{url}

\newcolumntype{A}{>{\centering\arraybackslash}p{2.6cm}}
\newcolumntype{B}{>{\centering\arraybackslash}p{1.2cm}}
\newcolumntype{C}{>{\centering\arraybackslash}p{1.5cm}}

\title{Inference and Visualization of Community Structure in Attributed Hypergraphs Using Mixed-Membership Stochastic Block Models}

\author[1, *]{Kazuki Nakajima}
\author[2]{Takeaki Uno}

\affil[1]{Graduate School of Systems Design, Tokyo Metropolitan University, 6-6 Asahigaoka, Hino-shi, 191-0065, Tokyo, Japan}
\affil[2]{Principles of Informatics Research Division, National Institute of Informatics, 2-1-2 Hitotsubashi, Chiyoda-ku, 101-8430, Tokyo, Japan}
\affil[*]{Corresponding author: nakajima@tmu.ac.jp}

\begin{document}
\date{}
\maketitle

\begin{abstract}
Hypergraphs represent complex systems involving interactions among more than two entities and allow the investigation of higher-order structure and dynamics in complex systems. Node attribute data, which often accompanies network data, can enhance the inference of community structure in complex systems. 
While mixed-membership stochastic block models have been employed to infer community structure in hypergraphs, they complicate the visualization and interpretation of inferred community structure by assuming that nodes may possess soft community memberships. 
In this study, we propose a framework, HyperNEO, that combines mixed-membership stochastic block models for hypergraphs with dimensionality reduction methods. Our approach generates a node layout that largely preserves the community memberships of nodes. We evaluate our framework on both synthetic and empirical hypergraphs with node attributes. We expect our framework will broaden the investigation and understanding of higher-order community structure in complex systems.
\end{abstract}


\section{Introduction}

A complex system is often represented as a network composed of nodes and pairwise interactions between the nodes.
Various mathematical and computational methods have been developed to investigate the structure and dynamics of networks \cite{boccaletti2006, newman2018}.
Conventional modeling using dyadic networks, in which each edge connects a pair of nodes, however, may not accurately encode higher-order interactions among nodes (i.e., interactions among three or more nodes) in real-world complex systems.
Examples of such higher-order interactions include group conversations in social networks \cite{stehle2011, mastrandrea2015}, co-authoring in collaboration networks \cite{newman2001, patania2017}, and many more \cite{battiston2020}.
Such complex systems can be represented as hypergraphs composed of nodes and hyperedges, where a hyperedge represents interaction among two or more nodes.
In recent years, there has been notable progress in the development of measurements, dynamical process models, and theories for hypergraphs \cite{battiston2020}.

Community detection is a fundamental task that aims to describe the structure of a network by dividing the nodes into communities (i.e., sets of nodes such that each set of nodes is densely inter-connected).
In fact, community structure has been observed in empirical networks across various domains \cite{newman2002, fortunato2016}.
In social networks, for example, analyses of community structure revealed the organization of divisions in a research institution \cite{girvan2002} and the linguistic split of the Belgian population in mobile phone communications \cite{blondel2008}.
Higher-order interactions among nodes can complicate the definition and detection of community structure in a network, which motivates the development of computational methods for efficiently detecting high-quality communities \cite{kaminski2019, ruggeri2023}.

Among numerous approaches for detecting communities in networks \cite{fortunato2010, fortunato2016}, we focus on statistical inference methods using stochastic block models.
A stochastic block model is a generative model for random graphs that assumes multiple communities underlying the network and latent variables controlling intra- and inter-community interactions between nodes \cite{paul1983}.
Stochastic block models have been deployed for inferring latent community structure in empirical networks \cite{karrer2011, lee2019}.
Standard models assume `hard' community memberships of the nodes in a network (i.e., each node belongs to one community). 
In contrast, mixed-membership stochastic block models (MMSBMs) assume `soft' community memberships of the nodes in a network (i.e., each node may belong to multiple communities with different extents of propensity) \cite{airoldi2008}.
MMSBMs have been deployed to infer latent community structure in dyadic networks \cite{ball2011, debacco2017, contisciani2020}. 
Recent studies have extended MMSBMs for dyadic networks to the case of hypergraphs \cite{contisciani2022, ruggeri2023}.
Moreover, a useful application of MMSBMs is constructing a ``learned representation'' of the given network (i.e., the expected frequency of interactions involving each node pair) from the set of inferred latent variables \cite{airoldi2008}.

Node attribute data accompanies network data in many real-world complex systems.
Examples of node attributes include the age, ethnicity, and gender of individuals in social networks \cite{newman2016, contisciani2020} and affiliations of authors in collaboration networks \cite{pan2021, nakajima2023gender}.
While node attributes do not always align with communities in a network \cite{peel2017}, node attribute data potentially enhances the learning of community structure in dyadic networks \cite{newman2016, chunaev2020}.
For example, the school age of students contributes to the inference of community structure in a school friendship network \cite{newman2016}.
Higher-order interaction data often comes with node attribute data in real-world complex systems \cite{chodrow2020}.
Based on these previous studies, a recent study has developed an MMSBM for incorporating node attribute data into learning community structure in hypergraphs \cite{badalyan2024}.

It is still complicated to visualize and interpret inference results on community structure for MMSBMs because MMSBMs assume nodes can have soft community memberships.
To address this technical issue, we employ dimensionality reduction methods. 
Dimensionality reduction methods have been used for visualizing high-dimensional data in various domains, such as single-cell transcriptomic data \cite{huang2022}. 
Moreover, some graph layout methods also employ dimensionality reduction methods \cite{harel2004,yang2014,kruiger2017}. 
For instance, Kruiger et al.~proposed a graph layout method that uses distance matrices for graph representations and a modified cost function in the t-SNE dimensionality reduction method \cite{kruiger2017}. 
In our study, we exploit learned hypergraph representations for MMSBMs to compute distance matrices for use in dimensionality reduction methods.

This study proposes a framework, {\it HyperNEO}, that combines MMSBMs for hypergraphs with dimensionality reduction methods.
We focus on two specific MMSBMs for hypergraphs: (i) Hy-MMSBM \cite{ruggeri2023} and (ii) HyCoSBM \cite{badalyan2024}.
Hy-MMSBM is an MMSBM designed for hypergraphs and efficiently infers latent community structure in empirical hypergraphs \cite{ruggeri2023}.
HyCoSBM is an extension of the Hy-MMSBM, adapted for hypergraphs with node attributes \cite{badalyan2024}.
In our framework, we apply a dimensionality reduction method to a learned representation of the given hypergraph to generate a node layout in a two-dimensional space.
We find that node layouts generated by our framework facilitate the visualization and interpretation of inferred community structure in hypergraphs.
The source code to reproduce our results is available at \url{https://github.com/kazuibasou/hyperneo}.
A preprint of this study is available \cite{nakajima2024}.

\section{Methods}

\subsection{Hypergraph with node attributes}

A hypergraph consists of a set of nodes $V = \{v_1, \ldots, v_N\}$ and a set of hyperedges $E$, where $N$ is the number of nodes.
Any hyperedge $e$ is a subset of $V$, and its size $|e|$ is two or larger.
We denote by $D$ the maximum size of the hyperedge in the hypergraph and $\Omega$ the set of all subsets of $V$ that have sizes of two or larger, including $V$ itself.
We represent the hypergraph by an adjacency vector $\mathcal{A} = (A_e)_{e \in \Omega}$, where $A_e$ represents the weight of hyperedge $e \in \Omega$.
We assume that $A_e$ is the number of times $e$ appears in given data for any $e \in \Omega$.
In addition, each node $v_i$ belong to one of the $Z$ categories, as represented by a $Z$-dimensional attribute vector $(x_{iz})_{1 \leq z \leq Z}$, where $x_{iz} = 1$ if node $v_i$ belong to the $z$-th category and $x_{iz} = 0$ otherwise for any $z \in \{1, \ldots, Z\}$. 
We represent the attribute vectors for all the nodes as an $N \times Z$ matrix, $\bm{X} = (x_{iz})_{1 \leq i \leq N,\ 1 \leq z \leq Z}$.

\subsection{Framework} \label{section:2.1}

\begin{figure*}[t]
  \begin{center}
	\includegraphics[width=1.0\textwidth]{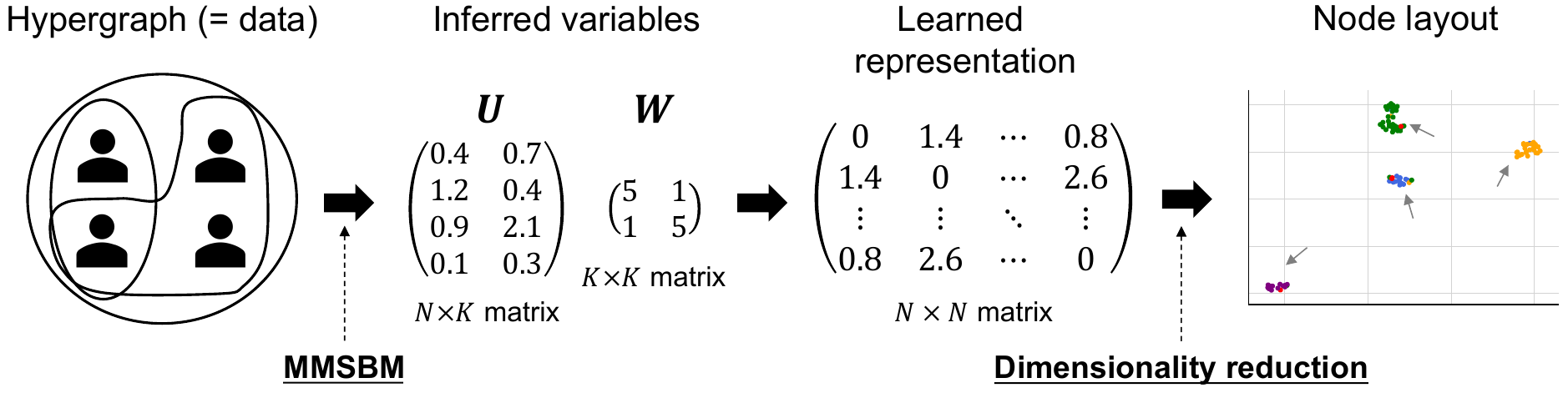}
  \end{center}
  \caption{Overview of the present framework. $N$: number of nodes. $K$: number of communities.}
  \label{fig:1}
\end{figure*}

In this section, we present a framework, {\it HyperNEO}, that combines MMSBMs for hypergraphs and dimensionality reduction methods.
Figure \ref{fig:1} shows its overview.

Suppose that a hypergraph data is given at hand.
We assume that the hypergraph accompanies node attribute data, although our framework can be applied to hypergraphs without node attributes.
Our framework also can be applied to graphs (i.e., hypergraphs in which the size of any hyperedge is two).
First, we fit the latent variables of an MMSBM to the hypergraph.
Our framework assumes MMSBMs that contain two latent variables: community membership matrix $\bm{U}$ and affinity matrix $\bm{W}$ (e.g., \cite{airoldi2008}).
We focus on Hy-MMSBM \cite{ruggeri2023} and HyCoSBM \cite{badalyan2024} as such MMSBMs for hypergraphs, although our framework is not restricted to these two MMSBMs.
Second, we construct a ``learned representation'' of the hypergraph, represented by an $N \times N$ matrix, based on the estimates of $\bm{U}$ and $\bm{W}$ and the generative model of the MMSBM.
Finally, we apply a dimensionality reduction method to the learned representation to place the $N$ nodes in a two-dimensional space.

In the following, we first briefly describe Hy-MMSBM and HyCoSBM.
Then, we describe the procedure for generating a node layout.

\subsubsection{Hy-MMSBM} \label{section:2.2.1}

Hy-MMSBM models a hypergraph probabilistically, assuming an underlying $K$ communities and soft community memberships of $N$ nodes \cite{ruggeri2023}.
The propensity of node $v_i$ belonging to each community is specified by a $K$-dimensional vector $(u_{ik})_{1 \leq k \leq K}$, where $u_{ik} \geq 0$ for any $i = 1, \ldots, N$ and any $k = 1, \ldots, K$.
We represent the membership vectors for all the nodes as an $N \times K$ matrix, denoted by $\bm{U} = (u_{ik})_{1 \leq i \leq N,\ 1 \leq k \leq K}$.
The strength of intra- and inter-community interactions among nodes is represented by a symmetric $K \times K$ affinity matrix, denoted by $\bm{W} = (w_{kq})_{1 \leq k \leq K,\ 1 \leq q \leq K}$, where $w_{kq} \geq 0$ for any $k = 1, \ldots, K$ and any $q = 1, \ldots, K$.
The latent variables of Hy-MMSBM is $\bm{\theta} = (\bm{U}, \bm{W})$.

Hy-MMSBM assumes that the weight of hyperedge $e$ follows the Poisson distribution given by \cite{ruggeri2023}
\begin{align}
P(A_e\ |\ \bm{U}, \bm{W}) &= e^{-\left(\frac{\lambda_e}{\kappa_{|e|}}\right)} \frac{\left(\frac{\lambda_e}{\kappa_{|e|}}\right)^{A_e}}{A_e!}, 
\label{eq:1}
\end{align}
where
\begin{align}
\lambda_e &= \sum_{v_i \in e} \sum_{v_j \in e \backslash \{v_i\}} \sum_{k=1}^K \sum_{q=1}^K u_{ik} u_{jq} w_{kq}, \label{eq:2} \\
\kappa_s &= \frac{s(s-1)}{2} \binom{N-2}{s-2}. \label{eq:3}
\end{align}
Hy-MMSBM also assumes that the weights of the hyperedges in $\Omega$ are conditionally independent given $\bm{U}$ and $\bm{W}$ \cite{ruggeri2023}:
\begin{align}
P(\mathcal{A}\ |\ \bm{U}, \bm{W}) = \prod_{e \in \Omega} P(A_e\ |\ \bm{U}, \bm{W}).
\label{eq:4}
\end{align}
Then, the log-likelihood of $\mathcal{A}$ is given by \cite{ruggeri2023}
\begin{align}
\mathcal{L}_{\mathcal{A}}(\bm{U}, \bm{W}) = \sum_{e \in E} A_e \log (\lambda_e) -C \sum_{i=1}^{N} \sum_{j=1,\ j \neq i}^N \sum_{k=1}^K \sum_{q=1}^K u_{ik} u_{jq} w_{kq}, 
\label{eq:5}
\end{align}
where
\begin{align}
C = \sum_{s=2}^D \frac{1}{\kappa_s} \binom{N-2}{s-2}
\label{eq:6}
\end{align}
and the terms not depending on $\bm{U}$ and $\bm{W}$ are discarded.
We describe the inference procedure for the latent variables of the Hy-MMSBM in Supplementary Section S1.

\subsubsection{HyCoSBM} \label{section:2.2.2}

HyCoSBM is an MMSBM for hypergraphs with node attributes \cite{badalyan2024}.
In addition to the two latent variables $\bm{U}$ and $\bm{W}$, HyCoSBM introduces a latent variable that controls the strength of the association between each community and each category of the node attribute.
This latent variable is represented by a $K \times Z$ matrix, $\bm{\beta} = (\beta_{kz})_{1 \leq k \leq K,\ 1 \leq z \leq Z}$, where $\beta_{kz} \geq 0$ for any $k = 1, \ldots, K$ and any $z = 1, \ldots, Z$. 
The latent variables of HyCoSBM is $\bm{\theta} = (\bm{U}, \bm{W}, \bm{\beta})$.
HyCoSBM assumes the following \cite{badalyan2024}: 
(i) $0 \leq u_{ik} \leq 1$ for any $i = 1, \ldots, N$ and any $k = 1, \ldots, K$; 
(ii) $\sum_{k=1}^K \beta_{kz} = 1$ for any $z = 1, \ldots, Z$;
(iii) $\mathcal{A}$ and $\bm{X}$ are conditionally independent given $\bm{\theta}$;
(iv) $\bm{U}$ and $\bm{W}$ probabilistically generate $\mathcal{A}$, whereas $\bm{U}$ and $\bm{\beta}$ probabilistically generate $\bm{X}$.

As with the Hy-MMSBM, the HyCoSBM assumes that the weight of hyperedge $e$ follows the same Poisson distribution given by Eq.~\eqref{eq:1} and also assumes that the weights of the hyperedges in $\Omega$ are conditionally independent given $\bm{U}$ and $\bm{W}$ \cite{badalyan2024}.
HyCoSBM assumes that the attribute of node $v_i$ follows a product of Bernoulli distributions given by \cite{badalyan2024}
\begin{align}
P(x_{i1}, \ldots, x_{iZ}\ |\ \bm{U}, \bm{\beta}) = \prod_{z=1}^Z \pi_{iz}^{x_{iz}} (1 - \pi_{iz})^{1 - x_{iz}},
\label{eq:7}
\end{align}
where 
\begin{align}
\pi_{iz} = \sum_{k=1}^K u_{ik} \beta_{kz}
\label{eq:8}
\end{align}
and it holds true that $0 \leq \pi_{iz} \leq 1$ for any $i = 1, \ldots, N$ and $z=1, \ldots, Z$ since it assumes $0 \leq u_{ik} \leq 1$ for any $i=1, \ldots, N$ and $k = 1, \ldots, K$, and $\sum_{k=1}^K \beta_{kz} = 1$ for any $z=1, \ldots, Z$.
HyCoSBM also assumes that the attributes of the nodes are conditionally independent given $\bm{U}$ and $\bm{\beta}$ \cite{badalyan2024}:
\begin{align}
P(\bm{X}\ |\ \bm{U}, \bm{\beta}) = \prod_{i=1}^N P(x_{i1}, \ldots, x_{iZ}\ |\ \bm{U}, \bm{\beta}).
\label{eq:9}
\end{align}
Then, the log-likelihood of $\bm{X}$ is given by \cite{badalyan2024}
\begin{align}
\mathcal{L}_{\bm{X}}(\bm{U}, \bm{\beta}) = \sum_{i=1}^N \sum_{z=1}^Z x_{iz} \log \sum_{k=1}^K u_{ik} \beta_{kz} + \sum_{i=1}^N \sum_{z=1}^Z (1 - x_{iz}) \log \sum_{k=1}^K (1 - u_{ik}) \beta_{kz},
\label{eq:10}
\end{align}
where the terms not depending on $\bm{U}$ and $\bm{\beta}$ are discarded.
We describe the inference procedure for the latent variables of the HyCoSBM in Supplementary Section S2.

\subsubsection{Node layout} \label{section:2.3}
Once we obtain an estimate of $\bm{\theta}$, we place the nodes in a two-dimensional space.
We first construct a learned representation, which is an $N \times N$ matrix denoted by $\bar{\bm{A}} = (\bar{a}_{ij})_{1 \leq i \leq N,\ 1 \leq j \leq N}$, as follows.
We define $\bar{a}_{ij}$ for each $i=1, \ldots, N$ and each $j = 1, \ldots, N$ such that $i \neq j$ as the expectation conditional on Eq.~\eqref{eq:1} of the sum of the weights over the hyperedges to which nodes $v_i$ and $v_j$ belong:
\begin{align}
\bar{a}_{ij} = \sum_{e \in E,\ v_i \in e,\ v_j \in e} \frac{\lambda_e}{\kappa_{|e|}}. 
\label{eq:11}
\end{align}
We calculate $\lambda_e$ for given hyperedge $e$ using the estimate of $\bm{\theta}$, according to Eq.~\eqref{eq:2}.
We define $\bar{a}_{ii} = 0$ for any $i = 1, \ldots, N$.

We then apply a dimensionality reduction method to the learned representation, $\bar{\bm{A}}$.
We focus on t-SNE \cite{laurens2008}, UMAP \cite{mcinnes2018}, TriMap \cite{amid2019}, and PaCMAP \cite{wang2021} as dimensionality reduction methods.
These methods have been employed for visualizing high-dimensional data across various domains (e.g., single-cell transcriptomic data \cite{huang2022}).
We use the `TSNE' function in the `scikit-learn' library \cite{sklearn}, the `UMAP' function in the `umap-learn' library \cite{mcinnes2018}, `TRIMAP' function in the `trimap' library \cite{amid2019}, and `PaCMAP' function in the `pacmap' library \cite{wang2021}.
We set the number of nearest neighbors (i.e., `perplexity' in the `TSNE' function, `n\_neighbors' in the `UMAP' function, `n\_inliers' in the `TRIMAP' function, and `n\_neighbors' in the `PaCMAP' function) as the average number of neighbors of the node, where node $v$ is a neighbor of node $w$ if $v$ and $w$ share at least one hyperedge.
We use the Euclidean distance as the distance metric in all the functions.
We use default settings for other parameters in both functions.
As recommended in Refs.~\cite{whiteley2022,huang2022}, we apply a principal component analysis to the learned representation, $\bar{\bm{A}}$, to reduce the number of dimensions to 30 for each node before applying any dimensionality reduction method.
Then, we apply a given dimensionality reduction method (i.e., t-SNE, UMAP, TriMap, or PaCMAP) to the reduced representation of $N \times 30$ matrix.

\section{Results} \label{section:3}

\subsection{Synthetic hypergraphs}

\begin{figure*}[p]
  \begin{center}
	\includegraphics[width=0.7\textwidth]{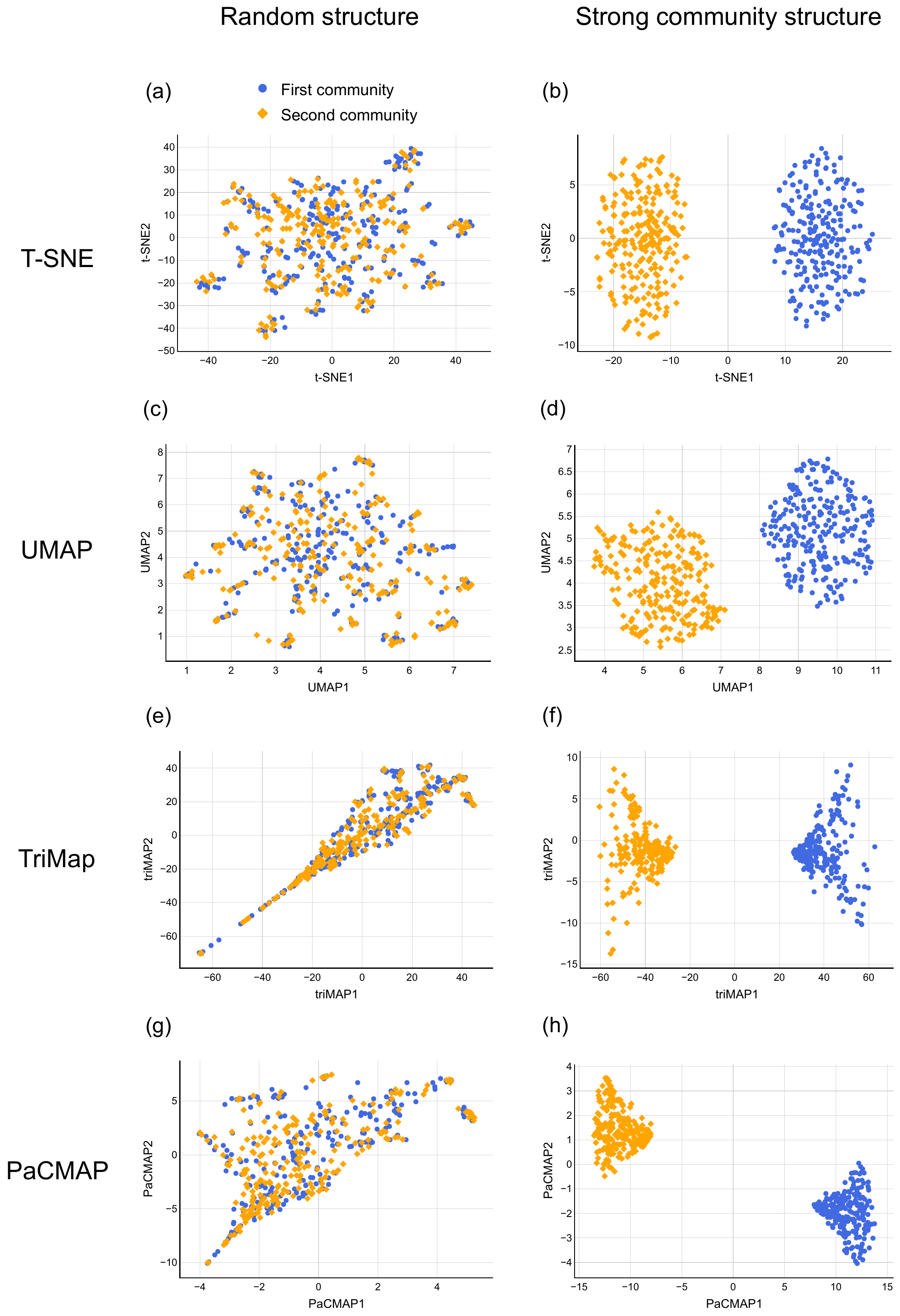}
  \end{center}
  \caption{Node layouts in synthetic hypergraphs with or without community structures. We set $(N, k, b) = (500, 3, 1)$ for generating all synthetic hypergraphs via the HSBM. We set $a=1$ in (a), (c), (e), and (g); we set $a = 10$ in (b), (d), (f), and (h); we show the results for t-SNE in (a) and (b), those for UMAP in (c) and (d), those for TriMap in (e) and (f), and those for PaCMAP in (g) and (h).}
  \label{fig:2}
\end{figure*}

We first apply our framework to synthetic hypergraphs without node attributes.
To generate synthetic hypergraphs, we use a stochastic block model for hypergraphs, HSBM \cite{kim2018}.
Consider a $k$-uniform hypergraph, where any hyperedge exactly contains $k$ nodes among $N$ nodes, with two equal-sized planted communities.
Let $V = \{v_1, \ldots, v_N\}$ be the set of nodes, and $\sigma \in \{-1, +1\}^N$ be an $n$-dimensional vector representing the community to which each node belongs.
Each hyperedge $e = \{v_{x_1}, \ldots, v_{x_k}\} \subseteq V$ appearing in the collection of all subsets with the cardinality $k$ of $V$ independently generates with the probability $p$ if and only if $\sigma_{x_1} = \sigma_{x_2} = \ldots = \sigma_{x_k}$, and with the probability $q$ otherwise, where $p$ and $q$ are parameters of the model.
We assume $p = a \log N / \binom{N-1}{k-1}$ and $q = b \log N / \binom{N-1}{k-1}$, where $a$ and $b$ are positive constants \cite{kim2018}.
We set $(N, k, b) = (500, 3, 1)$ and $a \in \{1, 10\}$.
Our framework employs Hy-MMSBM with the hyperparameter $K=2$, which is the original number of planted communities.

Figures \ref{fig:2}(a) and \ref{fig:2}(b) show the results for the case where we use t-SNE as a dimensionality reduction method.
The nodes are placed roughly randomly in a two-dimensional space for $a = b$ (see Fig.~\ref{fig:2}(a)).
This result is as expected because the HSBM for $a = b$ is an Erd{\H{o}}s-R{\'e}nyi hypergraph model, which does not produce community structure \cite{barthelemy2022}.
Theoretically, the two communities are correctly identified with a high probability if it holds true that $(\sqrt{a} - \sqrt{b})^2 / 2^{k-1} > 1$, as $N$ goes to infinity \cite{kim2018}.
This theoretical result should partially hold true for $a > 9$ in these synthetic hypergraphs since we set $(k, b) = (3, 1)$, although $N$ is small.
Indeed, for $a = 10$, our framework allows us to correctly identify the two communities in the two-dimensional space (see Fig.~\ref{fig:1}(b)).

We used three additional dimensionality reduction methods (i.e., UMAP \cite{mcinnes2018}, TriMap \cite{amid2019}, and PaCMAP \cite{wang2021}) in our framework.
We obtained qualitatively the same results for UMAP as those for t-SNE (see Figs.~\ref{fig:2}(c) and \ref{fig:2}(d)).
Figures \ref{fig:2}(e) and \ref{fig:2}(f) show the results for TriMap. 
Compared to the results for t-SNE and UMAP, while the coordinates of the nodes appear to lose some extent of randomness for $a=1$ (see Fig.~\ref{fig:2}(e), the nodes belonging to each community are densely positioned with each other for $a = 10$ (see Fig.~\ref{fig:2}(f)).
The results for PaCMAP are similar to those for TriMap (see Figs.~\ref{fig:2}(g) and \ref{fig:2}(h)).

\begin{figure*}[p]
  \begin{center}
	\includegraphics[width=0.7\textwidth]{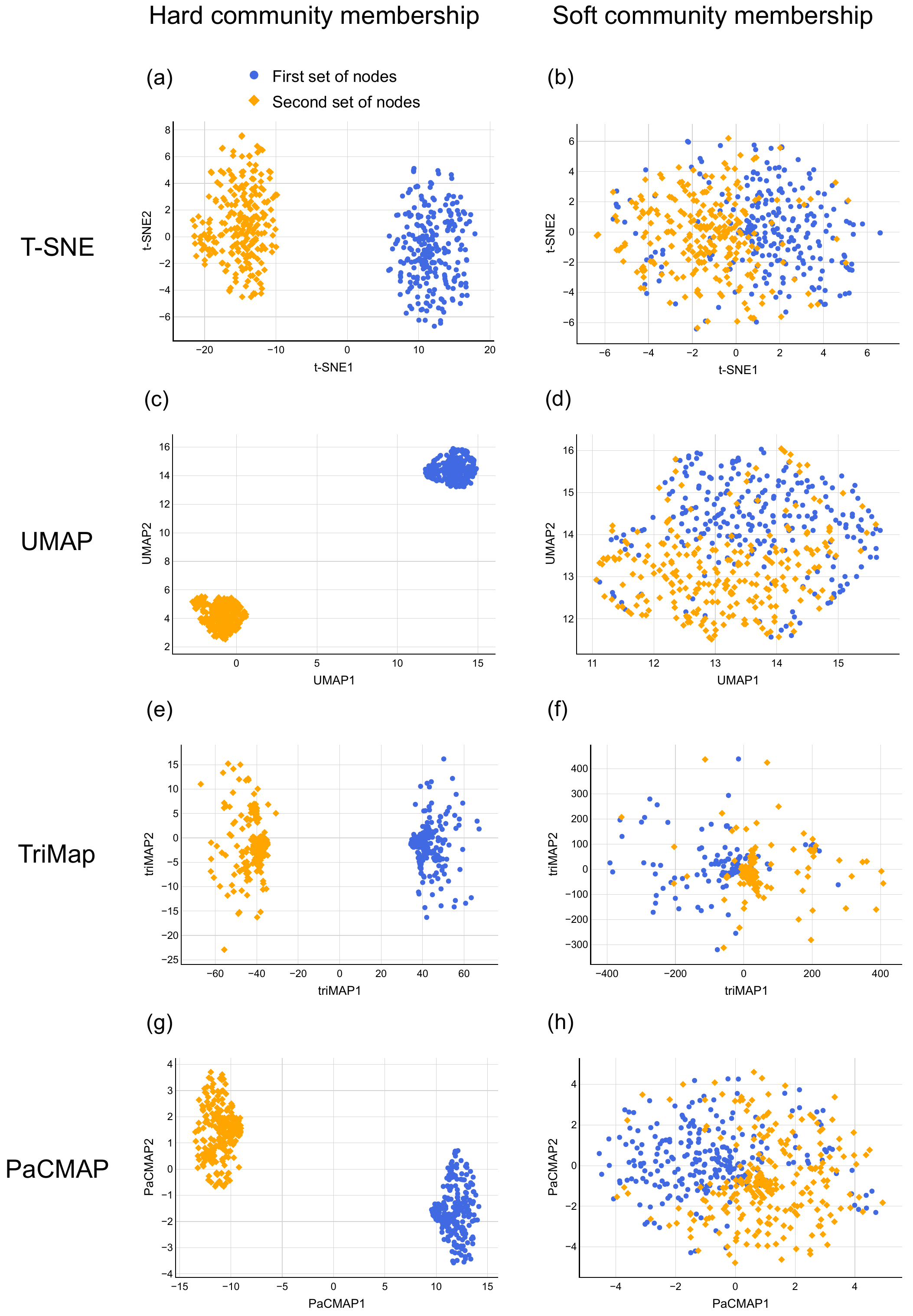}
  \end{center}
  \caption{Node layouts in synthetic hypergraphs with soft or hard community memberships. We set $(N, D, w_{\text{in}}) = (500, 3, 0.1)$ for generating all synthetic hypergraphs via the Hy-MMSBM. We set $\mu=1.0$ in (a), (c), (e), and (g); we set $\mu=0.8$ in (b), (d), (f), and (h); we show the results for t-SNE in (a) and (b), those for UMAP in (c) and (d), those for TriMap in (e) and (f), and those for PaCMAP in (g) and (h). We refer to the set of nodes that have the community membership $[\mu, 1-\mu]$ as `the first set of nodes' and the remaining nodes as `the second set of nodes'.}
  \label{fig:3}
\end{figure*}

For further assessment, we apply our framework to synthetic hypergraphs generated by a mixed-membership stochastic block model for hypergraphs, Hy-MMSBM \cite{ruggeri2023}.
Consider a hypergraph composed of $N$ nodes.
We define the community membership matrix $\bm{U}$ of the Hy-MMSBM as follows: we assign the membership vector $[\mu, 1-\mu]$ to $N/2$ nodes chosen uniformly at random and the membership vector $[1-\mu, \mu]$ to the remaining nodes, where we assume $0.5 \leq \mu \leq 1$.
Each node has a hard community membership for $\mu = 1$ and a soft community membership for $0.5 \leq \mu < 1$.
We define the affinity matrix $\bm{W}$ of the Hy-MMSBM by setting its diagonal entries to the value $w_{\text{in}} > 0$ and all other entries to 0.01.
Then, for each hyperedge $e$ appearing in the collection of all subsets with the cardinality $k \in \{2, \ldots, D\}$ of the set of nodes, we draw a random number $A_e$ from the Poisson distribution given by Eq.~\eqref{eq:1}.
We set $(N, D, w_{\text{in}}) = (500, 3, 0.1)$ and $\mu \in \{1.0, 0.8\}$.
Our framework employs Hy-MMSBM with the hyperparameter $K=2$.

Figure \ref{fig:3} shows the results for each dimensionality reduction method in our framework.
For any dimensionality reduction method, it is easier to identify the two sets of nodes in the two-dimensional space in the case of hard community memberships than in the case of soft community memberships, which is as expected.

\begin{figure*}[p]
  \begin{center}
	\includegraphics[width=0.7\textwidth]{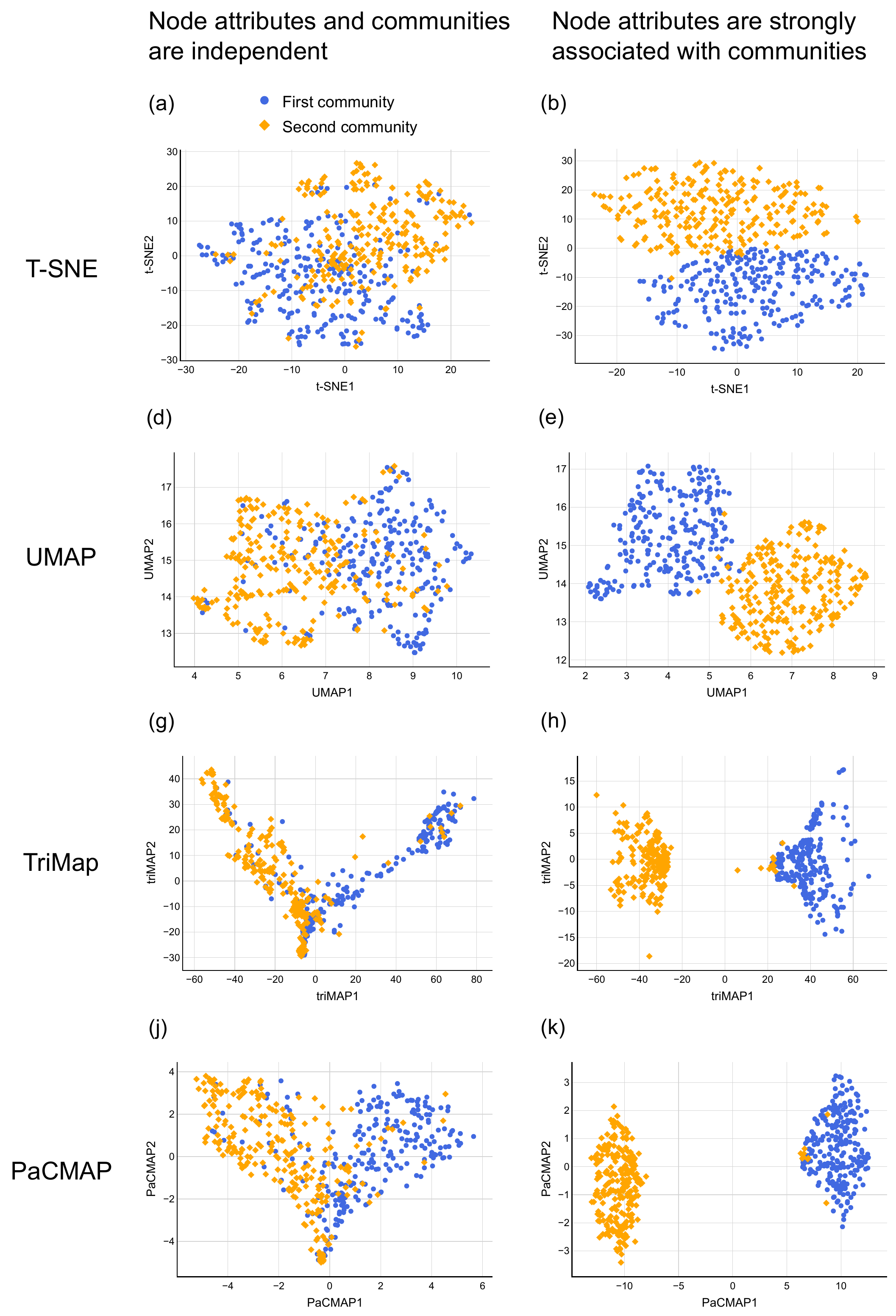}
  \end{center}
  \caption{Node layouts in synthetic hypergraphs with node attributes independent of or strongly associated with community structures. We set $(N, k, a, b) = (500, 3, 4, 1)$ for generating all synthetic hypergraphs via the HSBM. We set $r=0.5$ in (a), (c), (e), and (g); we set $r=0.9$ in (b), (d), (f), and (h); we show the results for t-SNE in (a) and (b), those for UMAP in (c) and (d), those for TriMap in (e) and (f), and those for PaCMAP in (g) and (h).}
  \label{fig:4}
\end{figure*}

To check the contribution of node attributes to the visualization of community structures, we apply our framework to synthetic hypergraphs with node attributes.
To this end, we generate a synthetic hypergraph using the HSBM for $(N, k, a, b) = (500, 3, 4, 1)$, which exhibits moderate community structure since $a > b$.
We assign the attribute of each node $v_i \in V$ as follows: we assign the true community label $\sigma_i$ with the probability $r \in [0.5, 1.0]$ as $v_i$'s attribute, and we randomly assign the label of $-1$ or $+1$ as $v_i$'s attribute otherwise.
The attribute of each node $v_i$ is independent of $v_i$'s community assignment for $r = 0.5$ and is completely consistent with it for $r=1.0$.
Our framework employs HyCoSBM with the hyperparameters $(K, \gamma)=(2, 2 r - 1)$ because it should be preferred not to incorporate node attributes at all ($\gamma \to 0$) for $r = 0.5$ and to completely incorporate ($\gamma \to 1$) for $r=1.0$ in the HyCoSBM.
We consider $r \in \{0.5, 0.9\}$.

Figure \ref{fig:4} shows the results for each dimensionality reduction method in our framework.
For any dimensionality reduction method, the nodes with the same community memberships are positioned more closely with each other in the case of $r=0.9$ (i.e., node attributes are strongly associated with community structure) than in the case of $r = 0.5$ (i.e., node attributes are independent of community structure), which is as expected.

To sum up, we find that node layouts generated by our framework facilitate the visualization of inferred community structure in synthetic hypergraphs.
If node attributes are associated with community structure, our framework using HyCoSBM allows us to visualize inferred community structure more effectively than the case where node attributes are uninformative.
Comparing the four dimensionality reduction methods, t-SNE and UMAP can successfully visualize randomized structures, and TriMap and PaCMAP can successfully visualize strong community structures.
In the following, unless we state otherwise, we use t-SNE as a dimensionality reduction method.

\subsection{Empirical hypergraphs}

\begin{table*}[t]
\caption{Properties of the empirical hypergraphs. $N$: number of nodes, $|E|$: number of hyperedges, $\bar{k}$: average degree of the node, $\bar{s}$: average size of the hyperedge, $D$: maximum size of the hyperedge, and $Z$: number of categories of the node attribute.}
\label{table:1}
\begin{center}
\begin{tabular}{| l | c c c c c c |} \hline
Data & $N$ & $|E|$ & $\bar{k}$ & $\bar{s}$ & $D$ & $Z$ \\ \hline
high-school & 327 & 7,818 & 55.6 & 2.3 & 5 & 9 \\
primary-school & 242 & 12,704 & 127.0 & 2.4 & 5 & 11 \\
hospital & 75 & 1,825 & 59.1 & 2.4 & 5 & 4 \\
workplace & 92 & 788 & 17.7 & 2.1 & 4 & 5 \\
house-committees & 1,290 & 335 & 9.2 & 35.3 & 81 & 2 \\
senate-committees & 282 & 301 & 18.8 & 17.6 & 31 & 2 \\
\hline
\end{tabular}
\end{center}
\end{table*}

\begin{table*}[t]
\caption{
Highest AUC in hyperedge prediction tasks and the corresponding hyperparameter set of each model in the empirical hypergraphs. We show the mean $\pm$ standard deviation of the AUC computed across $10^2$ independent train and test sets. We also show the $P$-value in the one-tailed Wilcoxon signed-rank test.
}
\label{table:2}
\begin{center}
\begin{tabular}{| l || A | B | A | B || C |} \hline
\multirow{2}{*}{Data} & \multicolumn{2}{c|}{Hy-MMSBM} & \multicolumn{2}{c||}{HyCoSBM} & \multirow{2}{*}{$P$-value} \\ 
& AUC & $K$ & AUC & $K, \gamma$ & \\ 
\hline
high-school & 0.834 $\pm$ 0.052 & 3 & 0.938 $\pm$ 0.007 & 9, 0.9 & $< 0.001$ \\ 
primary-school & 0.804 $\pm$ 0.050 & 3 & 0.911 $\pm$ 0.007 & 11, 0.8 & $< 0.001$ \\ 
hospital & 0.838 $\pm$ 0.018 & 2 & 0.840 $\pm$ 0.020 & 2, 0.3 & $0.09$ \\ 
workplace & 0.761 $\pm$ 0.035 & 5 & 0.821 $\pm$ 0.033 & 5, 0.9 & $< 0.001$ \\ 
house-committees & 0.934 $\pm$ 0.031 & 15 & 0.931 $\pm$ 0.031 & 4, 0.4 & 0.924 \\ 
senate-committees & 0.905 $\pm$ 0.043 & 3 & 0.903 $\pm$ 0.043 & 4, 0.7 & 0.939 \\ 
\hline
\end{tabular}
\end{center}
\end{table*}

In this section, we infer community structure in empirical hypergraphs with node attributes.
We use six datasets: high-school \cite{mastrandrea2015, chodrow2021, benson}, primary-school \cite{stehle2011, gemmetto2014, chodrow2021, benson}, hospital \cite{vanhems2013, ruggeri2023}, workplace \cite{genois2015, ruggeri2023}, house-committees \cite{house_committees}, and senate-committees \cite{senate_committees}.
See Table \ref{table:1} for properties of the empirical hypergraphs.
We first determine the hyperparameter set of each model in each empirical hypergraph based on the performance in hyperedge prediction tasks, as recommended in Refs.~\cite{contisciani2022, ruggeri2023}.
To this end, we compute the AUC in hyperedge prediction tasks for each candidate of the hyperparameter set in the Hy-MMSBM and HyCoSBM (see Supplementary Sections S1 and S2 for details). 

Table \ref{table:2} shows the highest AUC in hyperedge prediction tasks and the corresponding hyperparameter set of each model.
The AUC for the HyCoSBM is significantly higher than that for Hy-MMSBM in the workplace, high-school, and primary-school hypergraphs.
In contrast, the AUC for HyCoSBM is comparable with that for Hy-MMSBM in the hospital, house-committees, and senate-committees hypergraphs.
We also ran the one-tailed Wilcoxon signed-rank test using independent $10^2$ pairs of training and test sets of hyperedges, where the one-sided alternative hypothesis is that the AUC for Hy-MMSBM minus that for HyCoSBM is stochastically smaller than a distribution symmetric about zero.
These results suggest that node attribute data contributes to the inference of latent community structure in the first three hypergraphs, whereas little does in the hospital hypergraph.

As case studies, we explore associations between community structure and node attributes in the high-school and hospital hypergraphs.
We find that node attributes contribute to the inference of community structure in the high-school hypergraph, whereas they do little in the hospital hypergraph.
Unless we state otherwise, we use the tuned hyperparameter set that yields the highest AUC of each model in each hypergraph, shown in Table \ref{table:2}.

\begin{figure*}[t]
  \begin{center}
	\includegraphics[width=1.0\textwidth]{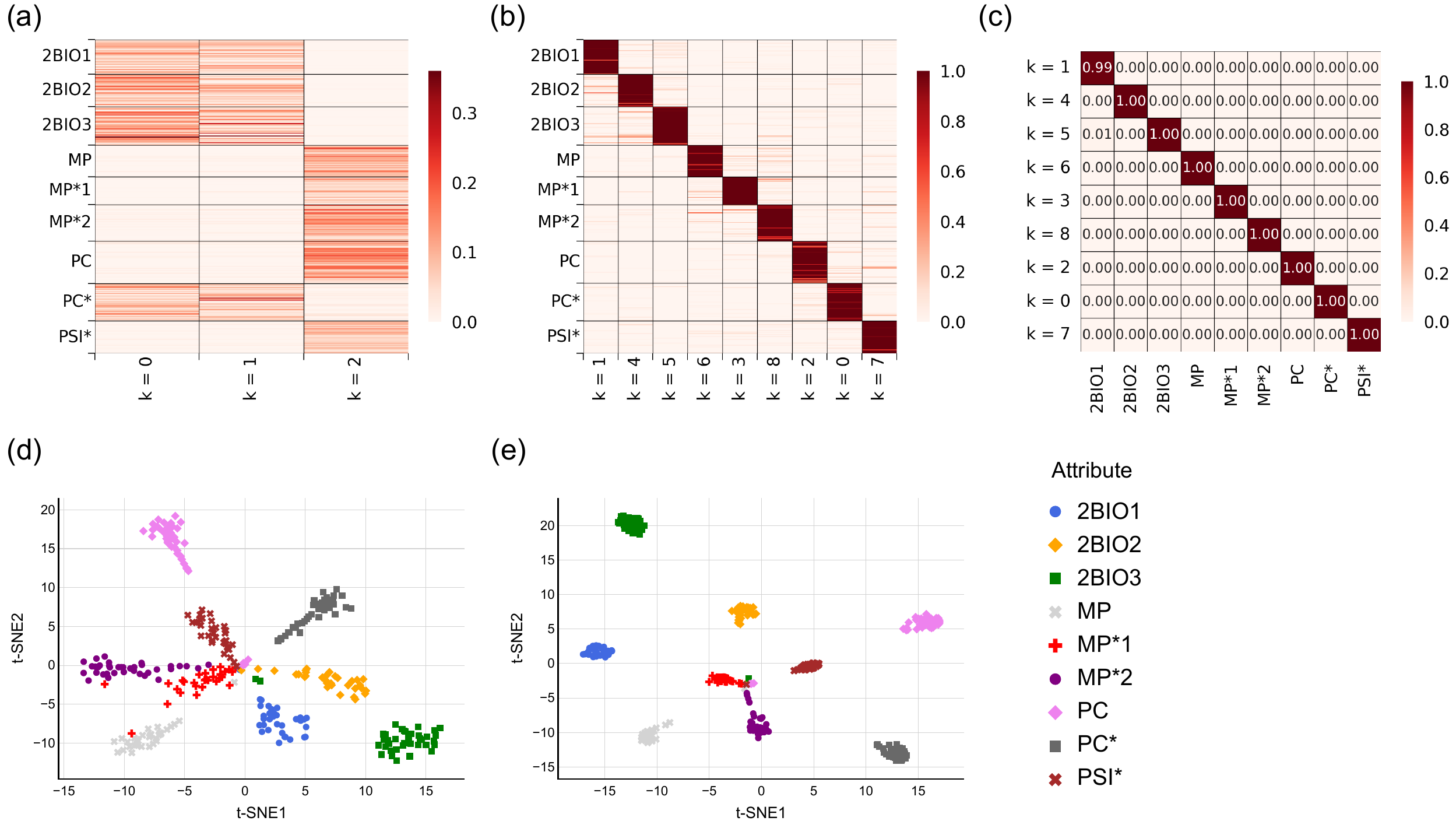}
  \end{center}
  \caption{Inference and visualization of community structure in the high-school hypergraph. (a) and (b): Inferred community membership matrices of (a) Hy-MMSBM and (b) HyCoSBM. (c): Inferred matrix $\bm{\beta}$ of HyCoSBM. (d) and (e): Node layouts generated using (d) Hy-MMSBM and (e) HyCoSBM. In panels (a) and (b), for visualization purposes, we arranged the $N$ row indices according to the attributes of the nodes, and we arranged the $K$ column indices arbitrarily.}
  \label{fig:5}
\end{figure*}

\begin{figure*}[t]
  \begin{center}
	\includegraphics[width=1.0\textwidth]{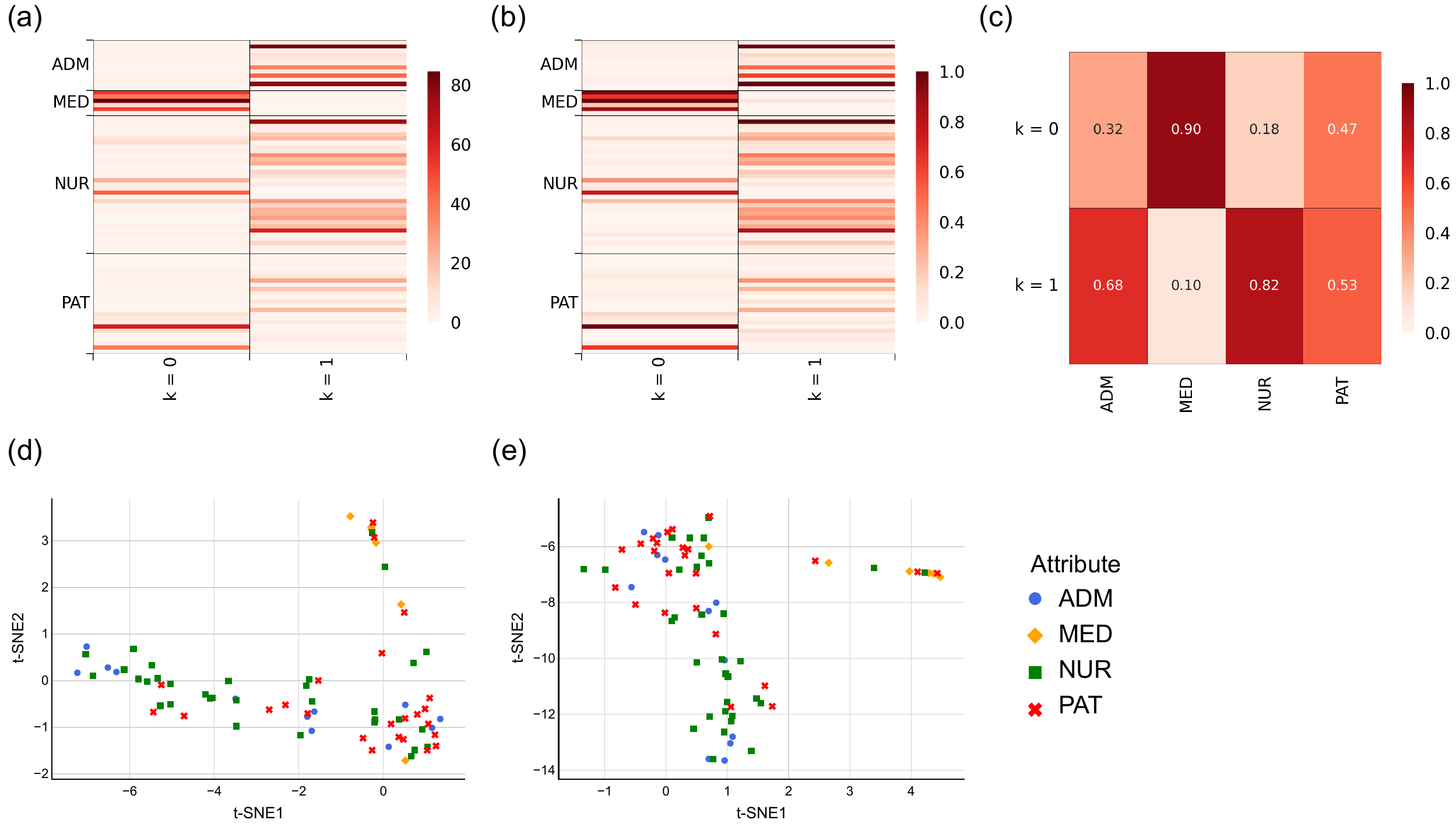}
  \end{center}
  \caption{Inference and visualization of community structure in the hospital hypergraph. (a) and (b): Inferred community membership matrices of (a) Hy-MMSBM and (b) HyCoSBM. (c): Inferred matrix $\bm{\beta}$ of HyCoSBM. (d) and (e): Node layouts generated using (d) Hy-MMSBM and (e) HyCoSBM. In panels (a) and (b), for visualization purposes, we arranged the $N$ row indices according to the attributes of the nodes, and we arranged the $K$ column indices arbitrarily.}
  \label{fig:6}
\end{figure*}

\subsubsection{High school}

We first focus on the high-school hypergraph composed of 327 students (i.e., nodes) in a high school in France and 7,818 contact events (i.e., hyperedges) among them \cite{mastrandrea2015, chodrow2021, benson}.
The attribute of a student is the class to which the student belongs.
Each student belongs to one of the nine classes: three classes on mathematics and physics (MP, $\text{MP}^*$1, and $\text{MP}^*$2), three classes on biology (2BIO1, 2BIO2, and 2BIO3), two classes on physics and chemistry (PC and $\text{PC}^*$), and one class on engineering studies ($\text{PSI}^*$).

Figures \ref{fig:5}(a) and \ref{fig:5}(b) show inferred community membership matrices of the Hy-MMSBM and HyCoSBM, respectively.
Note that there are no constraints on each entry $u_{ik}$ for the Hy-MMSBM, whereas a constraint (i.e., $0 \leq u_{ik} \leq 1$) for the HyCoSBM.
The result for the Hy-MMSBM suggests that students in the biology classes have similar community memberships, and mathematics and physics classes also do (see Fig.~\ref{fig:5}(a)).
The result for the HyCoSBM suggests that students have different community memberships in different classes (see Fig.~\ref{fig:5}(b)); this result is largely consistent with previous results in that students have more intra-class contacts than inter-class contacts \cite{mastrandrea2015}.
Figure \ref{fig:5}(c) shows an inferred matrix $\bm{\beta}$ of HyCoSBM, which represents the strength of the association between each community and each category of the node attribute.
The result suggests that each community is associated with a unique class of students.

Figures \ref{fig:5}(d) and \ref{fig:5}(e) show layouts of the students in a two-dimensional space using the Hy-MMSBM and HyCoSBM, respectively.
We find that students with similar community memberships are positioned closely together in the two-dimensional space for each model.
We also find that node layouts generated using HyCoSBM allow us to more easily identify communities of students in the two-dimensional space than those generated using Hy-MMSBM.

To sum up, we conclude that the high-school hypergraph has a community structure associated with the classes of students.
We also found that our framework generates layouts of the individuals into a two-dimensional space while largely preserving their community memberships.

\subsubsection{Hospital}

We next focus on the hospital hypergraph composed of 75 individuals (i.e., nodes) in a hospital and 1,825 contacts (i.e., hyperedges) among them \cite{vanhems2013, ruggeri2023}.
The attribute of an individual is the class to which they belong. 
Each individual belongs to one of the four classes according to their activity in the ward: patients (PAT), medical doctors (MED), paramedical staff (NUR), and administrative staff (ADM). 

The two MMSBMs yield qualitatively similar inferred community membership matrices (see Figs.~\ref{fig:6}(a) and \ref{fig:6}(b)). 
The results suggest that individuals in the same activity class do not strongly belong to the same community.
In contrast to the high-school hypergraph, the result for an inferred matrix $\bm{\beta}$ of HyCoSBM suggests that each community is associated with multiple activity classes (see Fig.~\ref{fig:6}(c)). 
Figures \ref{fig:6}(d) and \ref{fig:6}(e) show t-SNE layouts of the individuals generated using the Hy-MMSBM and HyCoSBM, respectively.
The individuals are relatively sparsely distributed over the space for each model.
On the other hand, for both models, the medical doctors are placed in roughly the same direction, and the administrative staffs are placed in a different direction from the medical doctors; these results largely align with the inferred community memberships of the individuals shown in Figs.~\ref{fig:6}(a) and \ref{fig:6}(b).

These results suggest no strong association between the activity classes of individuals and the community structure in the hospital hypergraph. 
In fact, the original time series data contain a high frequency of contacts between medical doctors and between paramedical staff but fewer contacts between patients and between administrative staff \cite{vanhems2013}. 
The limited number of contacts between patients may be due to the wards with mostly single rooms in the hospital \cite{vanhems2013}. 
Such contact patterns due to some localization of individuals (e.g., patients in different rooms may have few contacts) have also been discussed in Ref.~\cite{ueno2008}.

\section{Conclusions}

In this study, we introduced a framework, HyperNEO, that combines MMSBMs for hypergraphs with dimensionality reduction methods. 
Specifically, we employed the Hy-MMSBM and HyCoSBM as MMSBMs for hypergraphs, alongside t-SNE, UMAP, TriMap, and PaCMAP as dimensionality reduction methods.
Our framework generates node layouts that largely preserve community memberships of nodes.
We also confirmed that HyCoSBM improves the inference capability of community structure in hypergraphs compared to the Hy-MMSBM when node attributes are sufficiently associated with the communities, aligning with previous results \cite{badalyan2024}.
Our framework helps infer, visualize, and understand community structures in empirical hypergraphs with node attributes.
A future direction is to explore more effective combinations of stochastic block models and dimensionality reduction methods that allow accurate inference and informative visualization of higher-order community structures in real-world complex systems.

There are several limitations of this work. 
First, we may further investigate useful definitions of a learned representation, $\bar{\bm{A}}$, of the hypergraph in MMSBMs.
Theoretical analysis for the learned representation, including the spectral decomposition \cite{lei2015}, is a promising future work.
Second, we focused only on two MMSBMs for hypergraphs.
Our framework may extend to the case of stochastic block models with hard community memberships for hypergraphs (e.g., \cite{chodrow2021}).
Third, we applied our framework to only small synthetic and empirical hypergraphs.
Our framework may yield qualitatively different results on node layouts in large hypergraphs.
Fourth, we focused only on unordered attribute data (i.e., the index of the attribute category to which a node belongs is not informative).
It warrants future work to infer community structure in hypergraphs with node attributes that are ordered-discrete or continuous values \cite{newman2016, stanley2019}.

\subsection*{Acknowledgments}
We thank Naoki Masuda (State University of New York at Buffalo) for fruitful discussions.
KN thanks the financial support by the JST, ASPIRE Grant Number JPMJAP2328, JST, ACT-X Grant Number JPMJAX24CI, the Nakajima Foundation, and TMU local 5G research support.

\newpage

\begin{center}
\vspace*{12pt}
{\Large Supplementary material for:\\
\vspace{12pt}
Inference and Visualization of Community Structure in Attributed Hypergraphs Using Mixed-Membership Stochastic Block Models
}
\vspace{12pt} \\
\end{center}

\setcounter{figure}{0}
\setcounter{table}{0}
\setcounter{section}{0}
\setcounter{equation}{0}

\renewcommand{\thesection}{S\arabic{section}}
\renewcommand{\thefigure}{S\arabic{figure}}
\renewcommand{\thetable}{S\arabic{table}}
\renewcommand{\theequation}{S\arabic{equation}}

\begin{center}
\author{Kazuki Nakajima and Takeaki Uno}
\vspace{24pt} \\
\end{center}

\section{Hy-MMSBM}

\subsection{Inference} 

We aim to find $(\bm{U}, \bm{W})$ that maximizes $\mathcal{L}_{\mathcal{A}}(\bm{U}, \bm{W})$ using an expectation-maximization (EM) algorithm \cite{dempster1977}.
We first initialize independently $\bm{U}$ and $\bm{W}$ uniformly at random.
Then, we iterate alternating two optimization steps until a stopping criterion is met.
In one step, we update a variational parameter using the following equation \cite{ruggeri2023}:
\begin{align}
\rho_{ijkq}^{(e)} = \frac{u_{ik} u_{jq} w_{kq}}{\lambda_e},
\label{eq:7}
\end{align}
where $\rho_{ijkq}^{(e)}$ is a probability distribution that satisfies the condition $\sum_{v_i \in e} \sum_{v_j \in e \backslash \{v_i\}} \sum_{k=1}^K \sum_{q=1}^K \rho_{ijkq}^{(e)} = 1$.
In the other, we update $\bm{U}$ and $\bm{W}$ using the following equations \cite{ruggeri2023}:
\begin{align}
u_{ik} &= \frac{\sum_{e \in E,\ v_i \in e} A_e \sum_{j=1,\ j \in e,\ j \neq i}^N \sum_{q=1}^K \rho_{ijkq}^{(e)}}{C \sum_{j=1,\ j \neq i}^N \sum_{q=1}^K u_{jq} w_{kq}}, \label{eq:8} \\
w_{kq} &= \frac{\sum_{e \in E} A_e \sum_{i=1,\ v_i \in e}^{N} \sum_{j=1,\ v_j \in e,\ j \neq i}^N \rho_{ijkq}^{(e)}}{C \sum_{i=1}^{N} \sum_{j=1,\ j \neq i}^N u_{ik} u_{jq}}. \label{eq:9}
\end{align}

\subsection{Implementation} 
In our implementation, the iteration stops if at least one of the following conditions is met: (i) the number of iterations reaches $10^2$; (ii) the relative increase in the total log-likelihood per iteration is less than $10 ^{-3}$.
We observed that, in most cases, the relative increase in the total log-likelihood per iteration became less than $10 ^{-3}$ before reaching $10^2$ iterations.
Note that increasing the number of iterations too much can degrade inference results \cite{newman2016}.
In addition, to mitigate the issue that the EM algorithm is not guaranteed to converge to the global maximum, we generate $N_{\text{R}}$ independent initial configurations for $(\bm{U}, \bm{W})$, and we run the above inference procedure starting with each of them \cite{newman2016}.
Among the $N_{\text{R}}$ inferred results for $(\bm{U}, \bm{W})$, we choose the one that yields the highest final value of $\mathcal{L}_{\mathcal{A}}(\bm{U}, \bm{W})$.
We set $N_{\text{R}} = 10$.

\subsection{Determining $K$} 
To deploy Hy-MMSBM to a given hypergraph, we determine the hyperparameter $K$ value.
Unless we fix it a priori, we choose the $K$ value such that Hy-MMSBM achieves the highest accuracy in hyperedge prediction tasks on average among all candidates of $K$ value \cite{ruggeri2023}, as described below.

We perform hyperedge prediction tasks for a given $K$ value as follows. 
First, we sample 80\% of hyperedges in $E$ without replacement uniformly at random and regard them as a training set, $E_{\text{train}}$.
We regard the remaining hyperedges as a test set, $E_{\text{test}} = E\ \backslash\ E_{\text{train}}$.
Second, we input the training set $E_{\text{train}}$ and the hyperparameter $K$ to the Hy-MMSBM to infer the latent variables.
Third, we compute the area under the curve (AUC) for the prediction of the hyperedges in $E_{\text{test}}$ as follows.
For each hyperedge $e \in E_{\text{test}}$, we sample a hyperedge $e' \in \Omega\ \backslash\ E$ that satisfies the condition of $|e| = |e'|$ uniformly at random without replacement.
Let $R$ be the list of length $|E_{\text{test}}|$ of pairs $(e, e')$ thus obtained.
We define $\text{AUC} = \sum_{(e,\ e') \in R} f(e, e') / |R|$, where \cite{ruggeri2023}
\begin{align}
f(e, e') = 
\begin{cases}
1 & \text{if }P(A_{e}\ >\ 0)\ >\ P(A_{e'}\ >\ 0),\\
0.5 & \text{if }P(A_{e}\ >\ 0)\ =\ P(A_{e'}\ >\ 0),\\
0 & \text{otherwise}.
\end{cases}
\label{eq:10}
\end{align}
We compute $P(A_e > 0)$ for given hyperedge $e$ using the set of inferred latent variables, according to Eq.~(1).
We independently repeat the above process $10^2$ times to compute the mean and standard deviation of AUC for the given $K$ value.

We choose $K$ that yields the highest mean of AUC among all their candidates.
Unless we state otherwise, we examine $K \in \{2, 3, \ldots, 15\}$.

\section{HyCoSBM}

\subsection{Inference} 
We aim to find $(\bm{U}, \bm{W}, \bm{\beta})$ that maximizes the following log-likelihood function \cite{badalyan2024}:
\begin{align}
\mathcal{L}(\bm{U}, \bm{W}, \bm{\beta}) = (1 - \gamma) \mathcal{L}_{\mathcal{A}}(\bm{U}, \bm{W}) + \gamma \mathcal{L}_{\bm{X}}(\bm{U}, \bm{\beta}),
\label{eq:15}
\end{align}
where $\gamma \in [0,1]$ is a hyperparameter that controls the relative contributions of the structural and attribute terms.
To this end, we use an expectation-maximization (EM) algorithm.
We first initialize $\bm{\theta}$ uniformly at random.
Then, we iterate alternating two optimization steps until a stopping criterion is met.
In one step, we update the variational parameters using Eq.~\eqref{eq:7} and the following equations \cite{badalyan2024}:
\begin{align}
h_{izk} &= \frac{\beta_{kz} u_{ik}}{\sum_{k'=1}^K \beta_{k'z} u_{ik'}}, \label{eq:16} \\
h'_{izk} &= \frac{\beta_{kz} (1 - u_{ik})}{\sum_{k'=1}^K \beta_{k'z} (1 - u_{ik'})} \label{eq:17}.
\end{align}
In the other, we update $\bm{U}$, $\bm{W}$, and $\bm{\beta}$ as follows.
We take the smaller root of the following equation to update $u_{ik}$ \cite{badalyan2024}:
\begin{align}
\zeta_{ik} u_{ik}^2 - (\zeta_{ik} + \eta_{ik} + \xi_{ik}) u_{ik} + \eta_{ik} = 0, \label{eq:18}
\end{align}
where
\begin{align}
\zeta_{ik} &= C (1 - \gamma) \sum_{j=1,\ j \neq i}^N \sum_{q=1}^K u_{jq} w_{kq}, \label{eq:19} \\
\eta_{ik} &= (1-\gamma) \sum_{e \in E,\ v_i \in e} A_e \sum_{v_j \in e \backslash \{v_i\}} \sum_{q=1}^K \rho_{ijkq}^{(e)} + \gamma \sum_{z=1}^Z x_{iz} h_{izk}, \label{eq:20} \\
\xi_{ik} &= \gamma \sum_{z=1}^Z (1- x_{iz}) h'_{izk}. \label{eq:21}
\end{align}
We use the following equations to update $w_{kq}$ and $\beta_{kz}$ \cite{badalyan2024}:
\begin{align*}
w_{kq} = &\frac{\sum_{e \in E} A_e \sum_{i=1,\ v_i \in e}^{N} \sum_{j=1,\ v_j \in e,\ j \neq i}^N \rho_{ijkq}^{(e)}}{C \sum_{i=1}^{N} \sum_{j=1,\ j \neq i}^N u_{ik} u_{jq}}, \\
\beta_{kz} = &\frac{\sum_{i=1}^N \left[x_{iz} h_{izk} + (1 - x_{iz}) h'_{izk} \right]}{\sum_{i'=1}^N \sum_{z'=1}^Z x_{i'z'} h_{i'z'k}}.
\end{align*}

\subsection{Implementation} 

We use the same stopping criteria for the iteration as the Hy-MMSBM.
Specifically, the iteration stops if at least one of the following conditions is met: (i) the number of iterations reaches $10^2$; (ii) the relative increase in the total log-likelihood per iteration is less than $10 ^{-3}$.
We observed that, in most cases, the relative increase in the total log-likelihood per iteration became less than $10 ^{-3}$ before reaching $10^2$ iterations.
In addition, after generating $N_{\text{R}}$ independent initial configurations for $(\bm{U}, \bm{W}, \bm{\beta})$, we run the above inference procedure starting with each of them.
Among the $N_{\text{R}}$ inferred results for $(\bm{U}, \bm{W}, \bm{\beta})$, we choose the one that yields the highest final value of $\mathcal{L}_{\bm{X}}(\bm{U}, \bm{\beta})$.
We set $N_{\text{R}} = 10$.

\subsection{Determining $K$ and $\gamma$} 
Unless we fix them a priori, we choose the pair of $K$ and $\gamma$ values such that HyperNEO achieves the highest accuracy in hyperedge prediction tasks on average among all candidate pairs of $K$ and $\gamma$ values \cite{badalyan2024}.
We follow the same procedure to compute the AUC for the given $K$ and $\gamma$ values as the Hy-MMSBM.
Specifically, we first sample 80\% of hyperedges in $E$ without replacement uniformly at random and regard them as a training set, $E_{\text{train}}$.
We regard the remaining hyperedges as a test set, $E_{\text{test}} = E\ \backslash\ E_{\text{train}}$.
Second, we input the training set $E_{\text{train}}$ and the hyperparameters $K$ and $\gamma$ to the HyCoSBM.
Third, we compute the AUC for the prediction of the hyperedges in $E_{\text{test}}$ by $\text{AUC} = \sum_{(e,\ e') \in R} f(e, e') / |R|$.
We compute $P(A_e > 0)$ for given hyperedge $e$ using the set of inferred latent variables, according to Eq.~(1).
We independently repeat the above process $10^2$ times to compute the mean and standard deviation of AUC for the given pair of ($K$, $\gamma$).

We choose $(K, \gamma)$ that yields the highest mean of AUC among all these candidates.
Unless we state otherwise, we examine the $14 \times 9$ pairs of $K$ and $\gamma$, where $K \in \{2, 3, \ldots, 15\}$ and $\gamma \in \{0.1, 0.2, \ldots, 0.9\}$.

\renewcommand{\refname}{Supplementary References}

\end{document}